\documentclass[trackchanges]{aastex701}

\usepackage{xspace}
\usepackage{xcolor}

\newcommand{\logg}{\ensuremath{\log g}\xspace}
\newcommand{\teff}{\ensuremath{T_{\mathrm{eff}}}\xspace}
\newcommand{\feh}{\ensuremath{[\mathrm{Fe/H}]}\xspace}

\newcommand{\numax}{\ensuremath{\nu_{\rm{max}}}\xspace}
\newcommand{\dnu}{\ensuremath{\Delta\nu}\xspace}
\newcommand{\deltanu}{\dnu}
\newcommand{\muHZ}{\ensuremath{\mu}Hz\xspace}

\newcommand{\mstar}{\ensuremath{M_\star}\xspace}
\newcommand{\rstar}{\ensuremath{R_\star}\xspace}
\newcommand{\msun}{\ensuremath{\rm{M_\odot}}\xspace}
\newcommand{\rsun}{\ensuremath{\rm{R_\odot}}\xspace}

\begin{document}

\title{Detection of Solar-like Oscillations and Granulation at Infrared Wavelengths using JWST}

\author[orcid=0000-0000-0000-0001]{Angharad Weeks}
\affiliation{Sydney Institute for Astronomy, School of Physics, University of Sydney,
Sydney, NSW 2006, Australia}
\email[show]{angharad.weeks@sydney.edu.au}  

\author[orcid=0000-0001-8832-4488]{Daniel Huber} 
\affiliation{Institute for Astronomy, University of Hawai‘i,
2680 Woodlawn Drive, Honolulu, HI 96822, USA}
\email{huberd@hawaii.edu}

\author[orcid=0000-0001-5222-4661]{Timothy R. Bedding}
\affiliation{Sydney Institute for Astronomy, School of Physics, University of Sydney,
Sydney, NSW 2006, Australia}
\email{tim.bedding@sydney.edu.au}

\author{Jared Splinter}
\affiliation{Department of Earth $\&$ Planetary Sciences, McGill University, 3450 rue University, Montréal, QC, H3A 0E8, Canada }
\email{jared.splinter@mail.mcgill.ca}

\author[orcid=0000-0002-9037-0018]{Hans Kjeldsen}
\affiliation{Stellar Astrophysics Centre, Department of Physics and Astronomy, Aarhus University, Ny Munkegade 120, DK-8000 Aarhus C, Denmark }
\email{hans@phys.au.dk}

\author[]{Yaguang Li}
\affiliation{Institute for Astronomy, University of Hawai‘i,
2680 Woodlawn Drive, Honolulu, HI 96822, USA}
\email{yaguangl@hawaii.edu}

\author[]{Courtney Crawford}
\affiliation{Sydney Institute for Astronomy, School of Physics, University of Sydney,
Sydney, NSW 2006, Australia}
\email{courtney.crawford@sydney.edu.au;}

\author[]{Daniel Hey}
\affiliation{Institute for Astronomy, University of Hawai‘i,
2680 Woodlawn Drive, Honolulu, HI 96822, USA}
\email{danielhey@outlook.com}

\author{David Lafrenière}
\affiliation{Department of Earth $\&$ Planetary Sciences, McGill University, 3450 rue University, Montréal, QC, H3A 0E8, Canada }
\email{david.lafreniere@umontreal.ca}

\author{Louis-Phillipe Coulombe}
\affiliation{Planétarium de Montréal, Espace pour la Vie, 4801 av. Pierre-de Coubertin, Montréal, Canada}
\affiliation{Trottier Institute for Research on Exoplanets, Department of Physics, Université de Montréal, Montréal, Canada}
\email{louis-philippe.coulombe@umontreal.ca}

\author{Nicolas Cowan}
\affiliation{Department of Earth $\&$ Planetary Sciences, McGill University, 3450 rue University, Montréal, QC, H3A 0E8, Canada }
\affiliation{Department of Physics, McGill University, 3600 rue University, Montréal, QC, H3A 2T8, Canada}

\email{nicolas.cowan@mcgill.ca}

\begin{abstract}

Stellar oscillations and granulation are key probes of stellar convection, one of the most poorly understood processes in stars. Stellar models predict strong wavelength dependence of the amplitudes of oscillations and granulation, but hitherto both processes have only been measured at optical wavelengths. Here we present the first detection of oscillations and granulation at infrared wavelengths, using JWST time-series observations of the F6V main-sequence star WASP-121 that were obtained to study the atmosphere of an exoplanet. We find that the amplitudes of oscillations and granulation decrease with wavelength, consistent with expectations, but with a functional form that deviates from simple blackbody scaling. The detection opens the door to detailed tests of 3D convection models, facilitates the combined modeling of stellar and exoplanet signals in JWST passbands for exoplanet observations, and supports the predicted asteroseismic yields from the infrared time-domain data that will soon be obtained by the Roman Space Telescope.

\end{abstract}


\section{Introduction}

The surfaces of cool stars exhibit low-level variability due to processes arising from convection \citep[see][for a review]{Huber2025}. Granulation---the rising and falling plasma in convective cells---produces brightness fluctuations on timescales from minutes to hours. The granulation timescale, $\tau_{\rm{gran}}$, is related to the size of the granules, and has been shown to depend on stellar parameters \citep{trampedach1998}, namely surface gravity \citep{Mathur2011, Bastien2013}, \teff~\citep{KB2011,Kallinger2014, Yiren2020}, and metallicity \citep{Mathur2011, diaz2022}. 

As well as producing granulation, near‑surface convection also drives the stochastic excitation and damping of pressure‑mode (p‑mode) oscillations. These oscillations can be summarized by two global properties: the frequency of maximum power, \numax \citep{Brown1991}, and the average frequency separation between consecutive oscillation modes of the same angular degree, \dnu \citep{Tassoul1980}. These scale with stellar properties such as \logg, \teff and \feh \citep{KjeldsenBedding1995,KB2011}, and empirical scaling relations thus provide precise constraints on stellar density, radius, and age \citep{KjeldsenBedding1995,Bedding2014,Basu2016}. This has been demonstrated for hundreds of thousands of stars using space-based data from Kepler and TESS (The Transiting Exoplanet Survey Satellite, \citet{TESS}) \citep[see e.g.][]{Chaplin2013,Garcia2019, Garciastello2015,Jackiewics2021}.

\begin{figure}[ht!]
\plotone{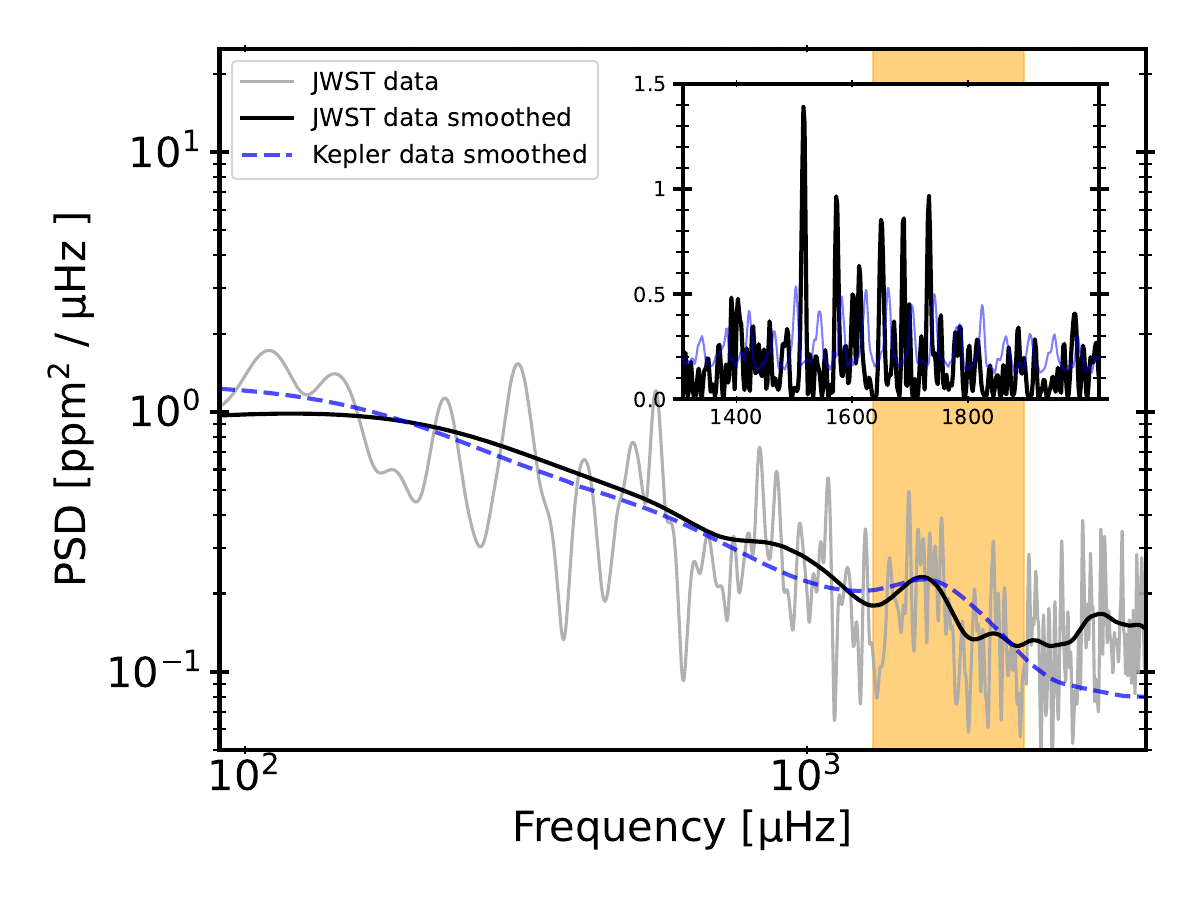}
\caption{Detection of solar-like oscillations and granulation in WASP-121 using JWST NIRISS/SOSS data.  PSD of the JWST NIRISS/SOSS Order 1 data for WASP-121 (grey and black), and the Kepler data for KIC 9139163 (blue dashed). The black line is a smoothed version of the data shown in grey, achieved by convolving with a Gaussian filter with width 4\dnu. The same smoothing was applied to the known solar-like oscillator KIC 9139163 \citep{lund2017}. Both smoothed curves exhibit a power excess, centred at approximately 1700~\muHZ. The shaded orange band shows the predicted \numax location for WASP-121, based on stellar parameters from \citet{Sing2024} and seismic scaling relations from \citet{KjeldsenBedding1995}. The inset shows the JWST data in linear frequency space, highlighting the oscillation envelope, which is found within the orange expectation band. Five evenly spaced peaks are visible. The smoothed Kepler power spectrum  is shown in blue.}

\label{fig:general}
\end{figure}

Amplitudes of stellar granulation and oscillations directly probe the physics of convective energy transport, which is poorly understood \citep{Houdek2006,HoudekDupret2015}. 
Traditional 1D models using Mixing Length Theory \citep[MLT,][]{bohm1958} and non-local MLT \citep[e.g.][]{Unno1967,Gough1977a,Gough1977b,Dupret2006} have been used to derive amplitude scaling relations based on energy equipartition \citep[e.g.][]{Goldreich1977b,Goldreich1977a,CD1983,CD1983b}, which show broad agreement with observations \citep{KjeldsenBedding1995, Huber2011,Kallinger2014}. 
Detailed 3D convection models of individual stars can predict amplitudes from first principles \citep[e.g.][]{Magic2013,Sulis2023,Samadi2013,RodruigezDiaz2022}, but are computationally expensive. Scaling relations derived from the results of 3D models offer the best test for the largest samples of stars, but still often require ad-hoc correction factors \citep[e.g.][]{Cranmner2014} to match observations. 

A powerful test of convection models would be the measurement of granulation and oscillation amplitudes at different wavelengths. 
Both scaling relations and 3D models predict that amplitudes of granulation and oscillations decrease at longer wavelengths \citep{KjeldsenBedding1995,Ballot2011, Lund2019, Samadi2013}, consistent with the assumption that convection can be approximated as small temperature fluctuations \citep{Michel2009, Ballot2011}.
However, this assumption has only been tested observationally over the small wavelength difference between Kepler and TESS bandpasses for stars other than the sun \citep[e.g.][]{Sreenivas2025}, or between the three visible filters from the SPM (Sun Photometers) on the VIRGO \citep[Variability of solar IRradiance and Gravity Oscillations,][]{virgo} mission,  for the Sun \citep{Michel2009,Ballot2011,lund2014a}.

The James Webb Space Telescope (JWST) offers the opportunity to study stellar variability due to convection at infrared wavelengths for the first time. JWST regularly obtains continuous light curves of stars to study exoplanet atmospheres that have excellent long-term stability and photometric precision \citep[see e.g.][]{Bean2018, Alderson23,Rustamkulov23,Feinstein23,lpc23,Bell24}. Granulation is expected to be a significant source of variability for some of these JWST observations (B. Morris $\&$ D. Huber, in prep), and has been reported as a likely source of correlated non-instrumental signal for some solar-type stars \citep{LPC2025, Splinter2025, teske2025, Hu2024, Boucher26}. However, no confirmed detection of oscillations and granulation has been made to date.

Here, we report the detection of solar-like oscillations and granulation in WASP-121 using 36.9 hours of NIRISS/SOSS \citep{Doyon2023, Albert2023} observations obtained as part of the NIRISS Exploration of the Atmospheric diversity of Transiting exoplanets (NEAT) Guaranteed Time Observations Programme (GTO 1201; PI Lafrenière). WASP-121 is a late F-type dwarf star, hosting an ultra-hot Jupiter on a 1.27 day orbit \citep{Delrez2016}. Scaling relations \citep{Brown1991}  predict a frequency of maximum power of $\numax$ $\simeq1730$ \muHZ, and smoothing the power spectrum, which is generated using the NIRISS/SOSS Order 1 integrated white light curve (centered at $\sim 1.5$ $\mu$m), reveals an oscillation excess that is consistent with this value. Comparison with a Kepler star with high SNR and similar stellar properties demonstrates a close match, further strengthening the detection. We also detected at least five evenly spaced peaks, which is a clear signature of solar-like oscillations.  This is the first detection of solar-like oscillations at infrared wavelengths in a star, including our own Sun.

\section{Asteroseismology} \label{sec:style}

We used the residuals of the NIRISS/SOSS Order 1 integrated lightcurve to search for signatures of stellar variability. We first converted the time series to a Lomb-Scargle periodogram, and then converted power into power spectral density (PSD). We used the asteroseismic scaling relations \cite{KjeldsenBedding1995, Brown1991} as follows, to predict the global seismic parameters, \numax and \deltanu:

\begin{equation}
    \nu_{max} \simeq \nu_{max}{}_{,\odot} \left(\frac{\mstar}{M_{\odot}}\right) 
    \left(\frac{\rstar}{R_{\odot}}\right)^{-2}
    \left(\frac{\teff}{\teff{}_{,\odot}}\right)^{-\frac{1}{2}} 
\end{equation}

\begin{equation}
    \dnu \simeq \dnu_{\odot} \left(\frac{M}{M_{\odot}}\right)^{\frac{1}{2}} \left(\frac{R}{R_{\odot}}\right)^{-\frac{3}{2}}.  
\end{equation}
Here $M_\star$, $R_\star$ and \teff are the stellar mass, radius and effective temperature adopted from \citep{Sing2024}, who used high precision spectroscopy, with values of $M_\star$ = $1.35 \pm 0.09$ \msun, $R_{\star}$ = $1.46 \pm 0.03$ \rsun, and \teff = $6628 \pm 140$K. This yielded a predicted  \dnu and \numax of $\sim 89$ $\mu$Hz and $\sim 1830$ $\mu$Hz, respectively.

Comparison to a known oscillator with similar physical parameters is a useful way to verify low-SNR detections of oscillations. We used a similar Kepler star, KIC 9139163 (HD 176071), which was part of the Kepler Legacy Sample \citep{lund2017,SA2017} -- one of the most precisely characterised samples of asteroseismic targets observed to-date. KIC 9139163 is similar to WASP-121 in \teff , mass, radius and \feh ($M_\star$ = $1.42 \pm 0.06$ \msun, $R_{\star}$ = $1.58 \pm 0.02$ \rsun, [Fe/H] = $0.15 \pm 0.09$, and \teff = $6400 \pm 84$ K), and has a measured \numax of $1730$ $\mu$Hz -- close to our predicted value for WASP-121.

The PSD of WASP-121 is shown in Fig. \ref{fig:general}. The solid black curve shows these data with a Gaussian filter of width 4\dnu, which confirms an oscillation envelope in the region of the predicted \numax, shown via the orange band. Applying the same smoothing to the KIC 9139163 data (the solid blue curve) confirms that the \numax is similar for both stars. The WASP-121 observations spanned only 36.9 hours (approximately 1.5 days), whereas the The KIC 9139163 data span a period of over 4 consecutive years, which explains the relatively low signal-to-noise ratio (SNR) of the WASP-121 detection.


The inset of Fig. \ref{fig:general} shows the WASP-121 power spectrum in linear frequency space, cropped around the predicted \numax. Five peaks are clearly visible around the between 1500 and 1900\muHZ, which are evenly spaced by a frequency value which is similar to the predicted \dnu. The PSD average over 15 1.5-day segments of the KIC 9139163 data is shown on the same plot in blue, for comparison.

\begin{figure}
    \centering
    \includegraphics[width=\linewidth]{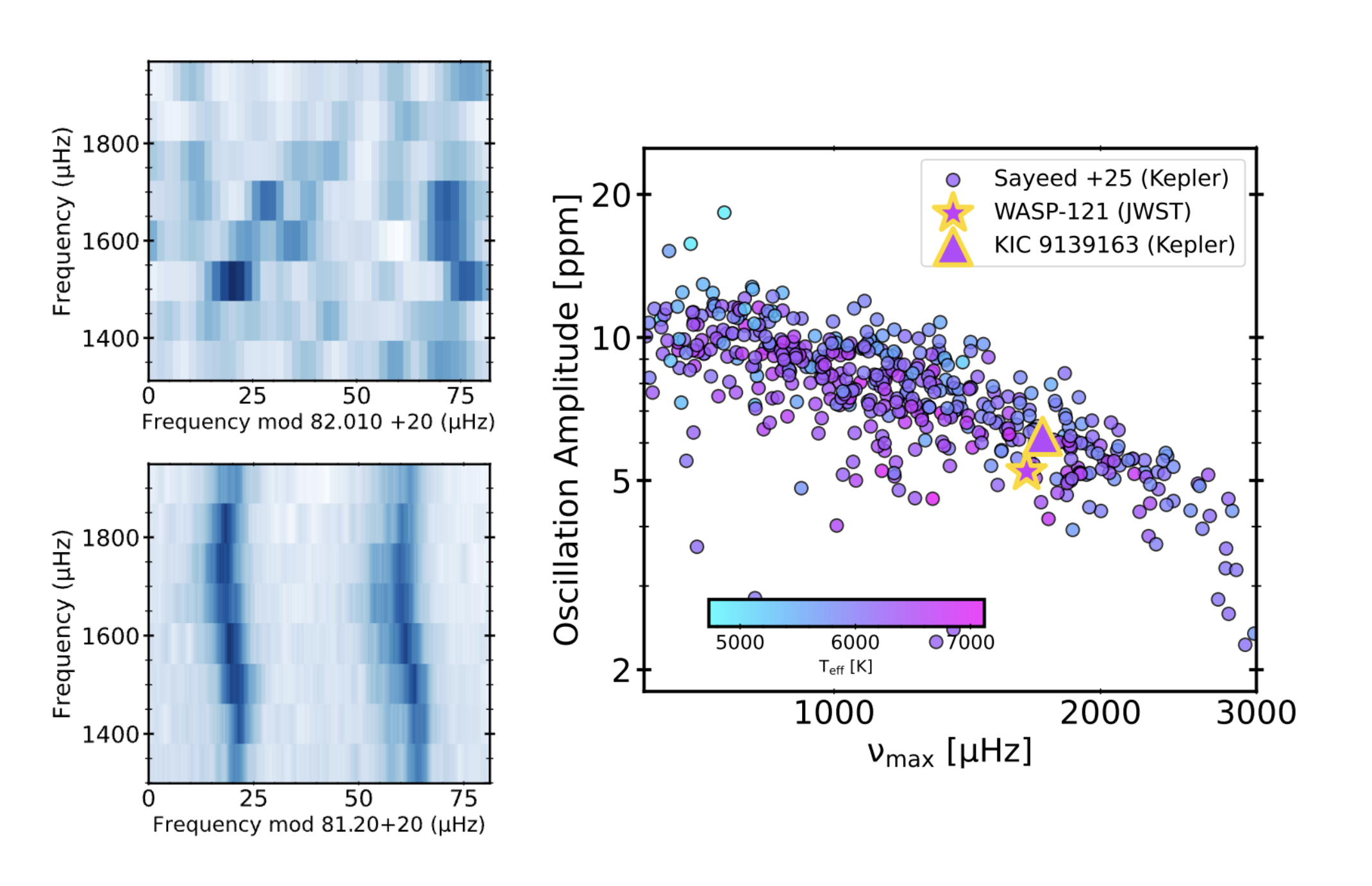}
    \caption{Left: Top: Échelle diagram of WASP-121. Bottom: Échelle diagram of KIC 9139163. Sections of the power spectrum, sliced at the value of \dnu are stacked. The vertical ridges confirm the detection of regularly spaced oscillations, which are in similar locations for both stars. The échelle for both stars is shifted left by 20 \muHZ to improve visibility. Right: Oscillation amplitude as a function of \numax for main-sequence Kepler stars from \citet{sayeed2025a}, colour-coded by \teff. WASP-121 and KIC 9139163 are plotted as the star and triangle markers.}
    \label{fig:fig2multipanel}
\end{figure}

To measure the \numax and \dnu, we used \texttt{pySYD} \citep{Chontos2022} to fit a Harvey-like background component and white noise component to remove the lower-frequency granulation signals. With this method, we measured a \numax of 1615 $\pm$ 67\muHZ with an amplitude of 5.21 $\pm$ 2.16 ppm.  The right hand panel of Fig. \ref{fig:fig2multipanel} compares these measurements to stars observed by Kepler \citep{sayeed2025a}, which show the well-known relation that stars with lower \numax\ oscillate with larger amplitudes. The amplitude of WASP-121 is slightly lower than the average amplitude of Kepler stars with similar \numax\ and \teff, such as with KIC 9139163 (A = 6.19 ppm $\pm$ 1.46 ppm), in agreement with expectations for infrared measurements.

We also compared the échelle diagrams of the two stars. The WASP-121 échelle shows only a few peaks, however the locations of the visible vertical ridges compare well to the échelle of the Kepler star, as seen in the right-hand panel of Fig. \ref{fig:fig2multipanel}. Mode identification is not possible with the JWST échelle alone, because hot F-type dwarfs typically exhibit blending between the l=0 and l=2 modes in frequency space, making mode identification challenging \citep[see e.g.][]{White2,white1}. However, by comparing to the well-resolved Kepler star, whose dipole and quadrupole modes are detectable individually, due to the 4-year Kepler dataset, we identified the l=0/2 and l=1 mode ridges on the left and right hand side of the WASP-121 échelle respectively.

\section{Granulation}
\label{sec:granulation}

Granulation, the surface signature of convection, introduces a broad low-frequency background in observations. We detect granulation in WASP-121, which appears as a slope with negative gradient at frequencies lower than the solar-like oscillations.
 Many studies use methods similar to \texttt{pySYD} to account for this `background', fitting one or more Harvey-like profiles \citep{Harvey1985} plus a white-noise term to the power spectrum \citep{Huber2011,Mathur2011,chontos21, Larsen2026}. However, frequency-domain fits can be unreliable for short or noisy datasets. We therefore  used a Gaussian Process (GP) to model oscillations and granulation simultaneously in the time domain. We found granulation amplitudes and frequencies consistent with predictions from literature \citep[e.g.][]{Kallinger2014}.

Similarly to \citet{LPC2025} and \citet{Splinter2025}, and following \citet{Foreman-Mackey2017}, we described the variability as the sum of two damped simple harmonic oscillator (\texttt{SHOTerm}) components implemented in \texttt{celerite2}: one for the granulation background and the other the p-mode oscillation envelope. The power spectral density of each term is
\begin{equation}
    S(\omega) = \sqrt{\frac{2}{\pi}}
    \frac{S_0\,\omega_0^4}{(\omega^2-\omega_0^2)^2 + \omega_0^2\omega^2 + \omega_0^2\omega^2/Q^2},
\end{equation}
where $S_0$ is the amplitude, $\omega_0$ the characteristic frequency, and $Q$ the quality factor.
 For the background component, we fix $Q=\frac{1}{\sqrt{2}}$  \citep{Foreman-Mackey2017, pereira2019}. For the oscillations, Q was allowed to vary. 
The full time-domain covariance of the kernels is 
\begin{equation}
    k(t_i,t_j) = k_{\mathrm{gran}}(t_i,t_j) + k_{\mathrm{osc}}(t_i,t_j).
\end{equation}

 We note that 
 the lower-frequency \textit{supergranulation} is challenging to measure due to the limited length of the dataset. We therefore recommend that with JWST data of similar length, only the median-timescale granulation is likely to be measurable with this method, especially with simultaneous detection of solar-like oscillations. A future paper will investigate the feasibility of measuring the granulation in datasets where oscillations are not detectable. A decline in the smoothed power spectrum of WASP-121 is indeed visible at several hundred $\mu$Hz (Fig.~\ref{fig:general}), consistent with the expected granulation frequencies \citep{Kallinger2014}. To robustly predict the frequency of granulation, we used the empirical scaling relation of \citet{Kallinger2014},
    \begin{equation}
        \nu_{\mathrm{gran}} = 0.317 \, \nu_{\max}^{0.970},
    \end{equation}
    together with our measured $\nu_{\max}$ of $1615$ $\mu\mathrm{Hz}$, yielding 
    $\nu_{\mathrm{gran}}\approx 439\,\mu\mathrm{Hz}$.

Using this estimate as a prior, we calculated the maximum a posteriori estimate of the GP model, and used the outputs as starting values to sample the likelihoods with \texttt{PyMC3}. For the white noise component, the mean $\mu_{w}$ was assigned a normal prior $\mathcal{N}(0,10)$, while its variance $S2_w$ followed a log-normal prior $\ln(\mathcal{N}(\overline{y_{\rm err}},10))$, centered on the mean reported measurement uncertainty. For the granulation component, the amplitude $A_{\rm{gran}}$ was given a log-normal prior $\ln(\mathcal{N}(\sigma_{y},50))$ based on the standard deviation of the light curve. We note that the generous prior width was chosen here such that the photon noise should not bias the measurement of the amplitude, which is further ensured by the simultaneous fitting of the white-noise term. The characteristic frequency $\nu_{\rm{gran}}$ has a normal prior $\mathcal{N}(439,50)~\mu\mathrm{Hz}$, and the quality factor $Q_{\rm{gran}}$ was fixed at $1/\sqrt{2}$. For the oscillation component, the amplitude $A_{\rm{osc}}$ followed a log-normal prior $\ln(\mathcal{N}(\sigma^2_{y},1))$, the frequency of maximum power $\nu_{\rm{max}}$ a normal prior $\mathcal{N}(1615,100)~\mu\mathrm{Hz}$, and the quality factor $Q_{\rm{osc}}$ a log-normal prior $\ln(\mathcal{N}(3,1))$. 

      


        

We applied the same GP approach to 30, 30-day sections of Kepler data for KIC 9139163, taking the mean and 1-sigma values of each GP parameter. Although granulation in Kepler stars is commonly modelled with a two-component mixture \citep{Kallinger2014}, doing so here would not provide a like-for-like comparison with the JWST analysis. The JWST time series has a model of a $\sim\!1.13$~day rotation period removed, using a semi-sinusoidal stellar activity model. We therefore applied a 2-term Savitzky-Golay filter \citep{1964AnaCh..36.1627S} to the Kepler data to remove variability on comparable timescales. After this pre-processing, we expected that the remaining \textit{Kepler} variability is dominated by granulation and the oscillation envelope.

\begin{figure*}
    \centering
    \includegraphics[width=\linewidth]{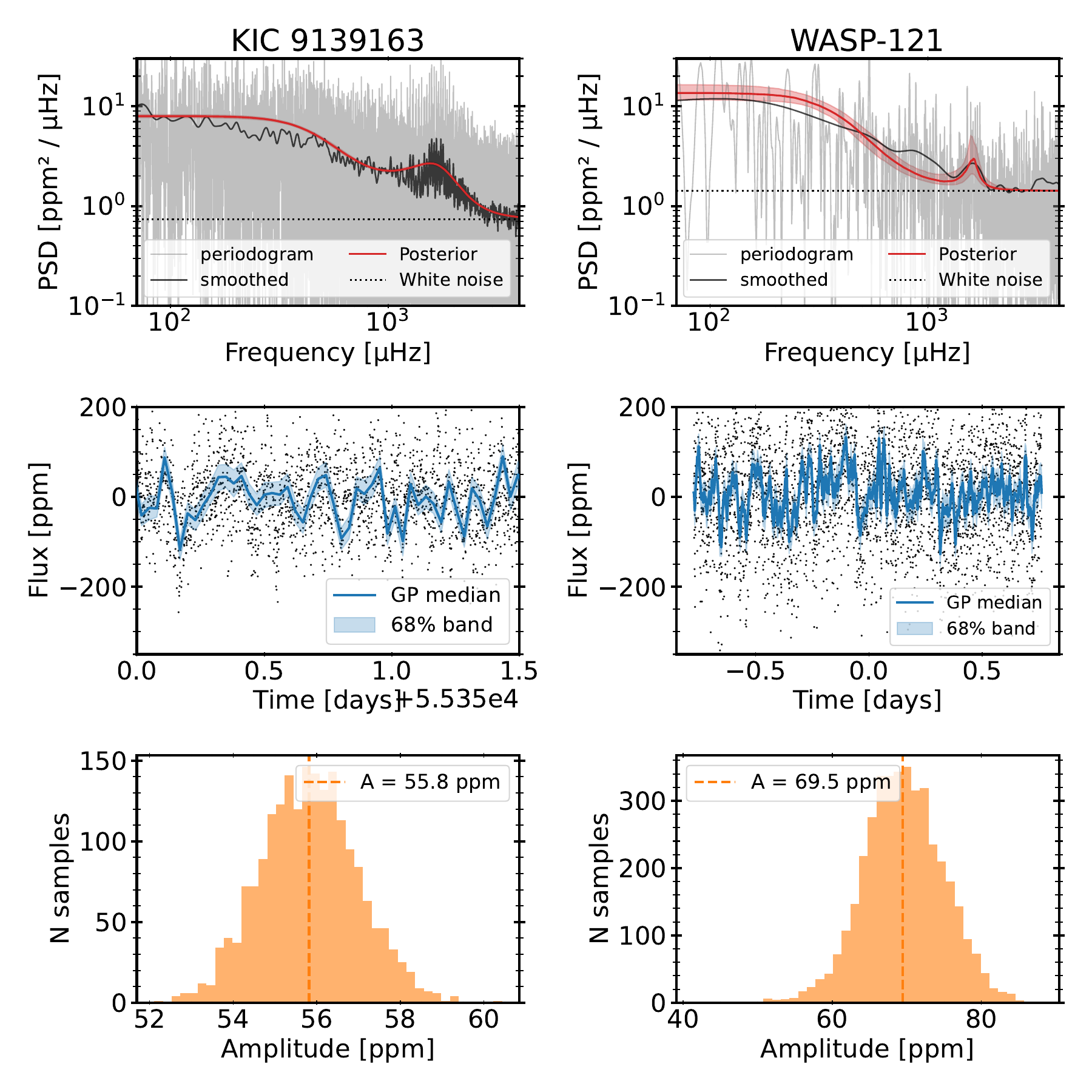}
    \caption{Granulation and oscillations are detectable with a two-component Gaussian Process applied to JWST NIRISS/SOSS Order 2 data of WASP-121. Gaussian process fits for WASP-121 (Right-hand panels), and one randomly selected 30-day section of KIC 9139163 (Right-hand panels). For both stars, the top panel shows the PSD manifestation of the GP, again with the median shown by the filled line, and the confidence interval in the shaded region. A clear granulation signal can be observed at a few hundred \muHZ~in each case, and an oscillation envelope model sits close to the observed \numax. The middle panels show the GP in the time domain, in blue, with the lighter shaded region showing the 68\% confidence interval. Only a ramdom 1.5 day excerpt of the Kepler data is shown, for comparison. The bottom panels show the posterior distributions for the amplitude of granulation in each star. }
    \label{fig:gp}
\end{figure*}

Similar granulation properties are determined for both stars. The GP in both the time and frequency domains is shown in Fig. \ref{fig:gp}, for both WASP-121 (right), and for one of the 30-day data sections of KIC 9139163 (left). In general, the amplitudes of the variability for both stars are similar, and the two-component GP model describes the convective processes well in both stars. For WASP-121, we measured $\numax$ = $1619^{+72.71}_{-78.22}~\mu\mathrm{Hz}$, $\dnu = 82.147 \pm 0.09~\mu\mathrm{Hz}$, granulation timescale $\tau_{\rm{gran}} = 9.79^{+1.26}_{-1.02}~\mathrm{min}$, and the granulation amplitude $A_{\rm{gran}} = 69.5 \pm 4.88~\mathrm{ppm}$. For KIC~9139163, the corresponding values were $\numax$ = $1686^{+68.82}_{-74.36}~\mu\mathrm{Hz}$, $\dnu = 81.2 \pm 0.09~\mu\mathrm{Hz}$, $\tau_{\rm{gran}} = 8.33^{+0.31}_{-0.28}~\mathrm{min}$, and $A_{\rm{gran}} = 60.5 \pm 2.28~\mathrm{ppm}$.

To validate the astrophysical nature of the oscillation and granulation signals, we considered the relationship between their amplitudes.This comparison was made without the correction from \citet{pereira2019}, such that the ratio was described by
\begin{equation}
    R = \frac{\sqrt{\frac{1}{\sqrt{2}}\omega_{0-gran} S_{0-gran} }}{\sqrt{\omega_{0-osc} S_{0-osc} Q_{osc}} }
\end{equation}. For the Order 1  NIRISS/SOSS data, this yielded a value of R=2.78. Using the equation and coefficients from Table 2 of \citet{Kallinger2014} gives a predicted R=3.10, which agrees well with the measured value.

\subsection{Testing the Chromaticity of Granulation}

The different observed amplitudes of granulation and oscillations in KIC 9139163 and WASP-121 may be attributed to the slight difference in stellar parameters, instrumental effects, or brightness. Therefore, these measurements alone are not sufficient to test for the wavelength dependence of the amplitudes. Such a test requires simultaneous data of the same star, made at different wavelengths.

\begin{figure}
    \centering
    \includegraphics[width=0.5\linewidth]{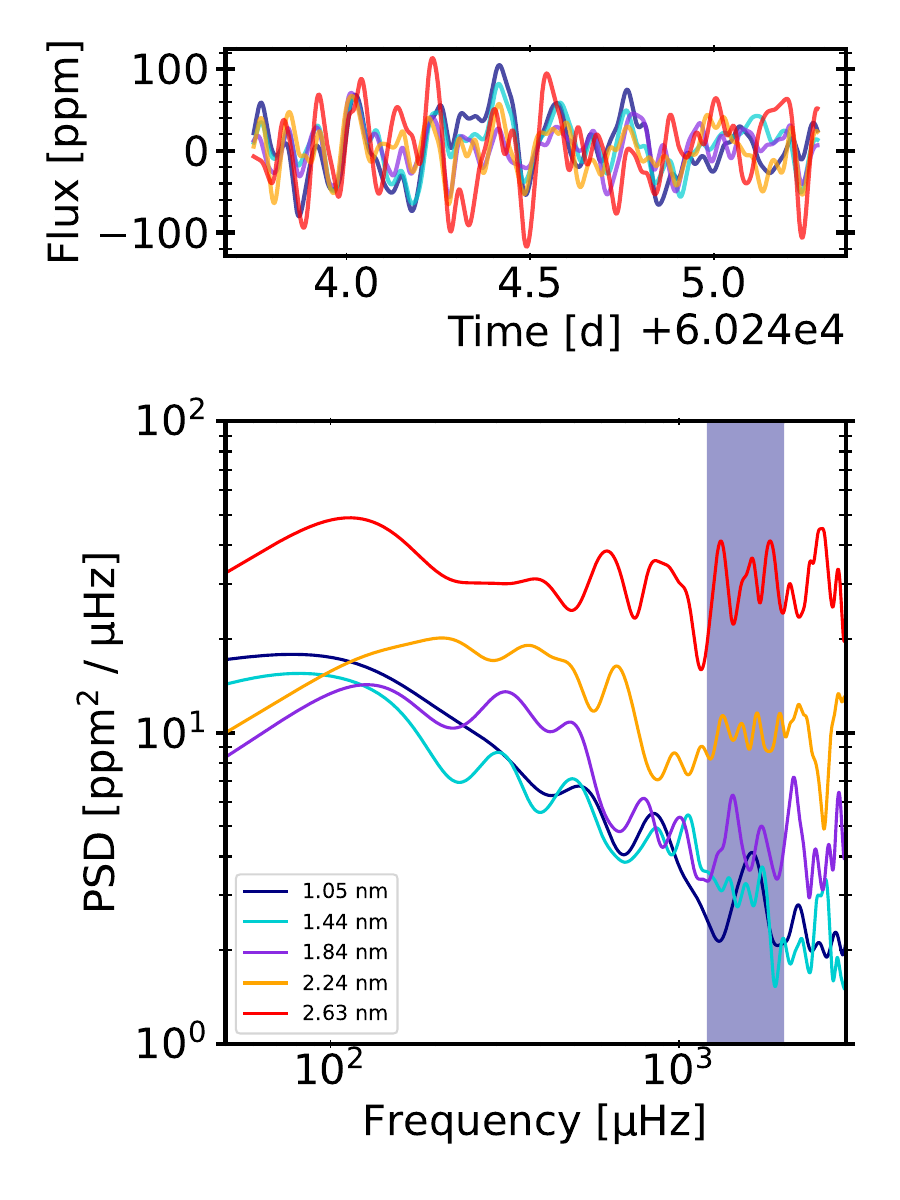}
    \caption{Top panel: Lightcurves of the 5 individual spectral time-series from Order 1, heavily smoothed with a gaussian convolution for visibility. Bottom panel: The same data converted to Power Spectral Density, with the predicted \numax highlighted by the blue band.}
    \label{fig:raw_binned_psds}
\end{figure}

To test whether the amplitudes of the signals observed vary with wavelength, we performed an identical analysis on the Order 2 data. Although Order 2 NIRISS/SOSS data is known to be of lower SNR \citep{Splinter2025}, oscillations were tentatively observed.

We repeated the GP analysis for Order 2 data, and 
found that the measured amplitude of granulation is lower in the Order 1 data than the  Order 2 data, with $A_{O1} = 71.9^{+4.9}_{-4.56}$ against $A_{O2} = 84.1^{+6.8}_{-5.8}$, in alignment with expectations from the literature \citep{Magic2013, Lund2019, Sreenivas2025}. These predict that the amplitudes of convective processes should decrease with wavelength, because the contrast between intergranular lanes and granule surfaces induces intensity differences set by the Planck function, which grows less sensitive to temperature at longer wavelengths.

The high signal-to-noise of the granulation allows us to measure its amplitude as a function of wavelength in multiple wavelength bins. To do this, we binned the 2048 individual pixel-level lightcurves, obtained as detailed in \citet{Splinter2025} (again without the GP model removed),
into 5 wavelength bins, centered at 1.05 $\mu$m, 1.44 $\mu$m, 1.84 $\mu$m, 2.24 $\mu$m, and 2.63 $\mu$m. The light curves and smoothed power spectra for each bin are shown in Fig. \ref{fig:raw_binned_psds}. We found that splitting the data into these 5 bins reduced the SNR, and therefore we could not statistically verify the presence of the oscillation power excess at all wavelengths. The excess can be seen visually in the first two bluest wavelength bins, however becomes indistinguishable from the white noise at redder wavelengths.

Therefore, we fit the GP with one kernel representing the granulation, fixing $\nu_{\rm{gran}}$ = 410 $\muHZ$ from the integrated Order 1 data, to avoid any shorter-timescale variation affecting the estimation of granulation timescale. We applied the GP methodology to the 5 binned spectral time series from Order 1, in addition to the white light Order 2 data, this time fixing frequency at the value measured from the integrated Order 1 data, and only allowing amplitude to vary.

\begin{figure*}
    \centering
    \includegraphics[width=\linewidth]{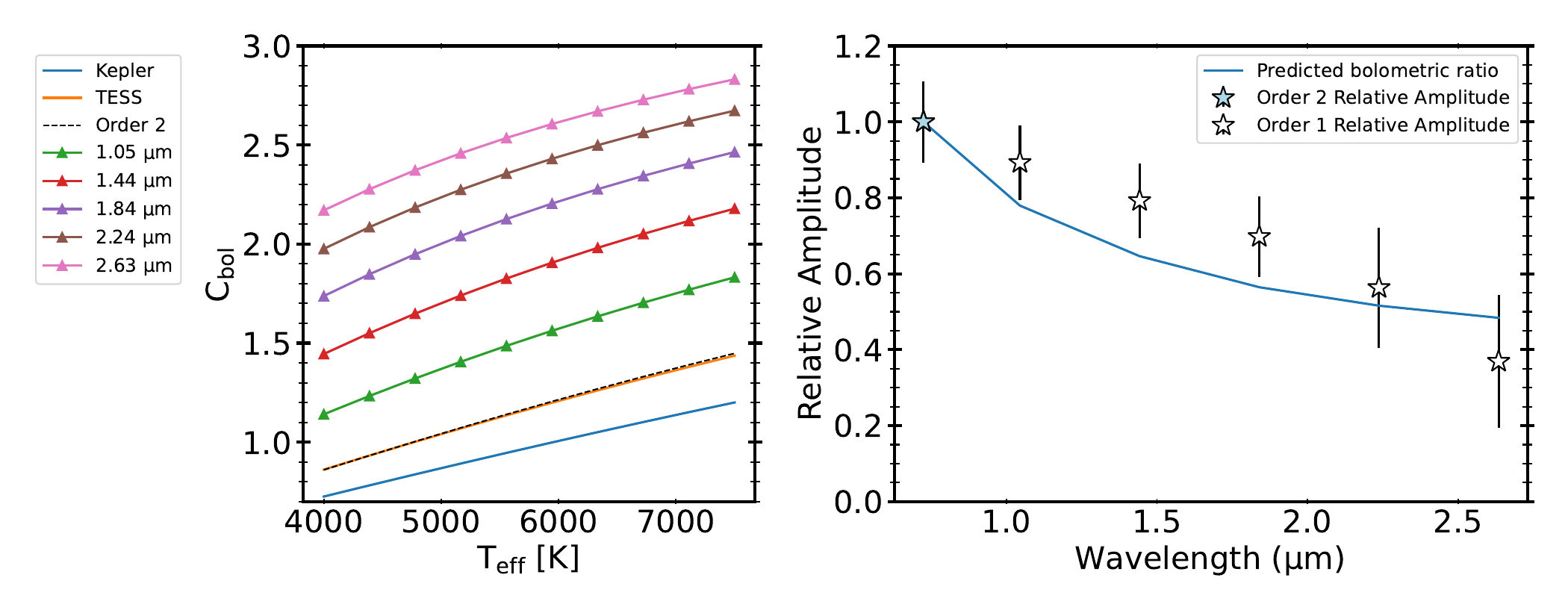}
    \caption{Left: Bolometric correction factors  ($C_{\rm{bol}}$) calculated with the formalism of \citep{Lund2019}, for 10 values of \teff, colour-coded by the wavelength range of the bandpass. Note that the values for TESS are almost identical to those for NIRISS/SOSS Order 1, as a result of their similar bandpasses. Right: The wavelength dependence of granulation measured with JWST. Open stars show the ratio of theogranulation amplitude of WASP-121 relative to the amplitude of granulation in the bluest wavelength bin. The blue line shows the predicted values assuming small temperature perturbations to a blackbody with an effective temperature of 6600 K, equivalent to WASP-121 (see text).}
    \label{fig:fig5multipanel}
\end{figure*}

Fig. \ref{fig:fig5multipanel} shows the GP granulation amplitude as a function of wavelength. The fifth and sixth bins yielded large error bars due to the higher RMS, so should be interpreted with caution. 
Nevertheless, a clear decrease in amplitude is observed with wavelength.

These measurements confirm that the amplitudes of granulation decrease with wavelength. In order to compare this result to predictions, we calculated expected relative amplitudes assuming a blackbody following the methodology from \citet{Lund2019} for each wavelength bin of the Order 1 data, and for the integrated  data.

In order to compare the observed amplitude decrease to theoretical predictions, we calculated the bolometric correction factor following \citet{Lund2019}. The bolometric amplitude of a signal must be divided by this factor to obtain the predicted amplitude at a given bandpass, such that 
\begin{equation}
    A_{bol}= C_{\rm{bol}}A_{\rm bp}
\end{equation}
where $A_{\rm{bol}}$ is the bolometric amplitude, and $A_{\rm{bp}}$  is the amplitude in a given bandpass. This bolometric factor allows us to build a simple wavelength-dependent model of the predicted amplitude dependence of a convective signal, like granulation.

If we model the atmosphere of a star as a blackbody (Planck) function, then it follows
\begin{equation}
F(\lambda,.) = B(\lambda,T_{\rm{eff}}).
\end{equation}
In flux density units this is given by
\begin{equation}
B(\lambda, T_{\rm{eff}}) = \frac{2\pi hc^2}{\lambda^5(e^{\frac{hc}{\lambda K_B T_{\rm{eff}}}}-1)}.
\end{equation}
The derivative of this with respect to $T_{\rm{eff}}$ is:
\begin{equation}
\frac{\partial{B(\lambda, T_{\rm{eff}})}}{\partial{T_{\rm{eff}}}} = \frac{2 \pi h^2 c^3 e^{\frac{hc}{\lambda K_B T_{\rm{eff}}}}}{\lambda^6 K_B T_{\rm{eff}}^2 (e^{\frac{hc}{\lambda K_B T_{\rm{eff}}}} -1)^2}.
\end{equation}
We use the above equations, along with the definition of a bolometric correction factor from \citet{Lund2019}, to define:
\begin{equation}
C_{bol} = \frac{4 \int \tau_p(\lambda) F(\lambda, .)d\lambda}{T_{\rm{eff}} \int \tau_p(\lambda) \frac{\partial{B(\lambda, T_{\rm{eff}})}}{\partial{T_{\rm{eff}}}}d\lambda}
\end{equation} Using the spectral response function for the NIRISS Order 2 spectral trace, and the \teff from \cite{Sing2024}, we calculate $C_{bol}$ for each wavelength bin used in our analysis. These values are used to plot the blue curve in the right panel of Fig. \ref{fig:fig5multipanel}. As a test of how the amplitude scaling affects the results, we exclude the Order 2 data and repeat the calculation and comparison for only the 5 Order 1 wavelength bins. This resulted in a slightly reduced deviation between measurement and prediction, however the decrease in amplitude with wavelength remained.

We also calculate values for the bolometric correction factor across the Order 1 wavelengths for a range of \teff between 4000 and 8000K. We present these factors in Table \ref{tab:bol} for use by the community, and plot them in the left-hand panel of Fig. \ref{fig:fig5multipanel}. The predictions made using this model, alongside the observations presented in this paper, reinforce the need to model the stellar variability independently in each spectral channel.

We find that the decrease in granulation amplitude broadly matches the predictions from blackbody models, which serves as observational evidence that the changes to stellar surfaces due to convective processes can be approximated as small adiabatic temperature fluctuations \citep{Samadi2013}. The data appear to follow a slightly different functional form to the predictions, with larger amplitudes at $\approx1.5\,\ \mu$m, possibly due to variability of spectral features due to granulation which have been predicted in 3D models \citep{Dravins23, witzke24, Sowmya26}.

We further validated the astrophysical nature of the stellar signals by measuring the correlation of total flux across wavelengths. A Pearson correlation test between consecutive spectral light curves produced correlation coefficients of 0.57, 0.38, 0.25 and 0.18, all with P-values $< 10^{-25}$. This confirms that the temporal parameter of the observed variability is, on average, consistent at all wavelengths. To isolate the granulation signals from oscillation signals, each time series was smoothed and high-pass filtered around a frequency of 1000\muHZ. We then fitted a model of the form $y=kx$ between the flux of the bluest band and each subsequent band. This resulted in values of K = $0.663 \pm 0.085$, $0.391 \pm 0.107$, $0.405\pm0.163$ and $0.337\pm0.279$. These provide simplified estimates for granulation amplitudes, which decrease with wavelength, in agreement with predictions and with our GP results. The difference in absolute values between this method and the GP fit can likely be attributed to the coarse signal isolation method, and the different treatment of white noise. 

\subsection{Implications} \label{subsec:tables}

The wavelength dependence of convection-driven variability has important implications across a range of astrophysics. Additional comparisons between observations in the infrared and theoretical scaling relations \citep{Samadi2013} will shed light on how well convection can be modeled as simple perturbations that only depend on temperature, and to what level 3D models accurately predict the time and wavelength dependence of granulation. Improved constraints on 3D models directly translate into 1D stellar structure model grids \citep{ludwig1999, Trampedach14, Amgic15, Jakob2018,Joyce2011}, which are widely used to study stellar populations within and outside of our galaxy.

The chromaticity of stellar variability is also important for transit spectroscopy studies. Unmodelled granulation can systematically bias planet radius measurements \citep{aigrain2004,Rackham23}, and the resulting error floor on planet radius can reach $10\%$, for an Earth-sized planet \citep{sulis2020a}. If planets transit on chords that have different brightnesses due to granulation, this can bias the interpretation of temporal atmospheric variations made from e.g. phase-curve analysis \citep{Chiavassa2017}. Incorrectly modelling these effects at multiple wavelengths could lead to wavelength-dependent biases in the results \citep{Rackham23,Fortune24} because transit depths measured at shorter wavelengths could suffer from more significant granulation noise floors than longer wavelengths \citep{Sarkar18}. Observations made at longer wavelengths are predicted to suffer from 10--20$\%$ transit depth error floors as a result of stellar variability (B. Morris $\&$ D. Huber, in prep). 

\begin{table}
\centering
\begin{tabular}{c c c c c c c}

\textbf{\teff [K]} &
\textbf{c$_{1.05}$} &
\textbf{c$_{1.44}$} &
\textbf{c$_{1.84}$} &
\textbf{c$_{1.975}$} &
\textbf{c$_{2.17}$} &
\textbf{c$_{0.86}$} \\
\hline
 4000 & 1.1408 & 1.4451 & 1.7376 & 1.9758 & 2.1702 & 0.8600 \\
 4444 & 1.2331 & 1.5502 & 1.8475 & 2.0853 & 2.2768 & 0.9323 \\
 4889 & 1.3215 & 1.6484 & 1.9483 & 2.1843 & 2.3723 & 1.0032 \\
 5333 & 1.4057 & 1.7402 & 2.0408 & 2.2741 & 2.4581 & 1.0724 \\
 5778 & 1.4859 & 1.8258 & 2.1259 & 2.3558 & 2.5356 & 1.1399 \\
 6222 & 1.5623 & 1.9059 & 2.2044 & 2.4305 & 2.6058 & 1.2055 \\
 6667 & 1.6349 & 1.9809 & 2.2769 & 2.4989 & 2.6698 & 1.2691 \\
 7111 & 1.7039 & 2.0510 & 2.3440 & 2.5617 & 2.7283 & 1.3308 \\
 7556 & 1.7696 & 2.1168 & 2.4063 & 2.6196 & 2.7820 & 1.3905 \\
 8000 & 1.8321 & 2.1786 & 2.4642 & 2.6732 & 2.8314 & 1.4483 \\
\hline

\end{tabular}

\caption{Bolometric corrections for JWST NIRISS/SOSS Order 1 wavelengths, under the Blackbody assumption.}
\label{tab:bol}
\end{table}

Finally, the wavelength dependence of oscillations is critical for estimating asteroseismic yields of future space-based missions. For example, the Nancy Grace Roman Telescope \citep{roman} will obtain continuous near-infrared time-domain observations in the Galactic Bulge, which is expected to detect oscillations in hundreds of thousands of red giants \citep{Weiss2025, Huber2023}. The results presented here support the assumptions made to predict those yields, and thus open up the opportunity to analyze variations in asteroseismic yields due to astrophysical effects such as stellar activity and close binaries \citep{Gaulme2014,Gaulme2016,Gaulme2019}.

\section{Conclusion}

In this paper, we used JWST/NIRISS SOSS observations of an exoplanet host star to presented the first detections of stellar granulation and solar‑like oscillations at infrared wavelengths. By modelling the time‑series data with a Gaussian‑process framework, we have estimated granulation timescale and oscillation envelope parameters for WASP‑121, which match predictions well. These results demonstrate the suitability of JWST for resolving near‑surface variability processes in the infrared. We also demonstrate, for the first time using observations, the chromaticity of convective processes in sun-like stars. This finding is crucial for our understanding of 3D stellar structure, mitigating stellar variability in multi-wavelength exoplanet observations, and for predicting the yields of future infrared asteroseismology surveys.

\begin{acknowledgments}
This work was supported by the Australian Research Council (DP 250102562). 
D.H. acknowledges support from the National Aeronautics and Space Administration (80NSSC25K7155, 80NSSC25K0177) and NASA’S Interdisciplinary Consortia for Astrobiology Research (NNH19ZDA001N-ICAR) under award number 19-ICAR19 2-0041. 
DL would like to acknowledge financial support from the Canadian Space Agency.
Some/all of the data presented
in this article were obtained from the Mikulski Archive
for Space Telescopes (MAST) at the Space Telescope
Science Institute. The specific observations analyzed
can be accessed via \dataset[doi:10.17909]{10.17909/ppjp-9743}. We acknowledge and appreciate the comments provided by the anonymous reviewer, which strengthened the quality of the manuscript. 
\end{acknowledgments}

\facilities{JWST (NIRISS)}

\software{ The work presented in this letter has made use of Astropy, \citep{astropy},
Matplotlib \citep{mpl}, NumPy \citep{numpy}, pandas \citep{pandas1,pandas2}, and Celerite \citep{Foreman-Mackey2017}.}

\bibliography{sample701}{}
\bibliographystyle{aasjournalv7}

\end{document}